\documentclass[a4paper,fleqn,usenatbib]{mnras}

\usepackage[T1]{fontenc}
\usepackage{ae,aecompl}

\usepackage{graphicx}	% Including figure files
\usepackage{amsmath}	% Advanced maths commands
\usepackage{amssymb}	% Extra maths symbols

\title[SMC star clusters]{Towards a comprehensive knowledge of the star cluster population 
in the Small Magellanic Cloud}

\author[Andr\'es E. Piatti]{
A.E. Piatti$^{1,2}$\thanks{E-mail: andres@oac.unc.edu.ar}
\\
$^{1}$Consejo Nacional de Investigaciones Cient\'{\i}ficas y T\'ecnicas, Av. Rivadavia 1917, 
C1033AAJ, Buenos Aires, Argentina\\
$^{2}$Observatorio Astron\'omico, Universidad Nacional de C\'ordoba, Laprida 854, 5000, 
C\'ordoba, Argentina\\
}

\date{Accepted XXX. Received YYY; in original form ZZZ}

\pubyear{2017}

\begin{document}
\label{firstpage}
\pagerange{\pageref{firstpage}--\pageref{lastpage}}
\maketitle

\begin{abstract}
The Small Magellanic Cloud (SMC) has recently been found to
harbour more than two hundred per cent increase of its known cluster population.
We provide here with solid evidence that such an unprecedented number of clusters
could be largely overestimated. On the one hand, the fully-automatic procedure
used to identify such an enormous cluster candidate sample did not recover
 $\sim$ 50 per cent, in average, of the known relatively bright clusters 
located in the SMC main body. On the other hand, the number of new
cluster candidates per time unit as a function of time results noticeably
different to the intrinsic SMC cluster frequency (CF), which should not be
the case if these new detections were genuine physical systems.
We additionally found that the SMC CF varies spatially,
in such a way that it resembles an outside-in process coupled with the
effects of a relatively recent interaction with the Large Magellanic Cloud.
By assuming that clusters and field stars share the same formation history,
we showed for the first time that the cluster dissolution rate also
depends on the position in the galaxy. The cluster dissolution
results higher as the concentration
of galaxy mass increases or external tidal forces are present.
\end{abstract}

\begin{keywords}
techniques: photometric -- galaxies: individual: SMC -- galaxies: star clusters:
general 
\end{keywords}

%%%%%%%%%%%%%%%%%%%%%%%%%%%%%%%%%%%%%%%%%%%%%%%%%%

%%%%%%%%%%%%%%%%% BODY OF PAPER %%%%%%%%%%%%%%%%%%

\section{Introduction}

The number of star clusters identified in the Small Magellanic Cloud (SMC) has steadily grown 
from a handful of objects firstly recognised by \citet{sw25} up to nearly six hundred clusters 
compiled in the catalogue of \citet[][hereafter B08]{betal08}.
Recently, \citet[][hereafter B18]{bitsakisetal2018} reported a list of 1108 clusters
that have not been reported before, distributed throughout the main body of the SMC.
They searched for stellar overdensities  on archival images of the 
Galaxy Evolution Explorer \citep[GALEX/NUV,][]{simonsetal2014}, the Swift Ultraviolet-Optical
Telescope (UVOT) Magellanic Clouds Survey \citep[SUMAC,][]{siegeetal2014}, the Magellanic Cloud 
Photometric Survey \citep[MCPS,][]{zetal02} and the "Surveying the
Agents of a Galaxy's Evolution SMC survey" \citep[SAGE-SMC,][]{gordonetal2011}, respectively. 
Then, they fitted theoretical isochrones from a Bayesian approach on field star cleaned
colour-magnitude diagrams (CMDs) of the cluster candidates. The whole process - from
the cluster search until the age estimate - was fully-automatically carried out. 

These new cluster candidates represent an increase of 215 per cent in the number
of known clusters spread within the same areal coverage in the B08's catalogue, which strikes 
our previous knowledge of the size of the SMC cluster population. Moreover, it is a bit of surprise 
that
such a huge number of new objects have been detected from images that do reach in average
comparable limiting magnitudes than those obtained by the  Optical Gravitational Lensing Experiment 
\citep[OGLE,][]{pieetal98}, which were used to perform the most recent 
update in the B08's  catalogue. 
%If we considered that the Large Magellanc Cloud (LMC) has a total 
%stellar mass $\sim$ 
%5 - 10 times bigger than that of the SMC \citep{beslaetal2012,vdmk14,yb2014}, the number
%of new detections would also challenge the proportion of SMC clusters to those catalogued
%in the LMC ($\sim 3000$, B08). 
Additionally, the spatial 
distribution of these new
detections also differentiates from that coming from B08. The former is more
broadly distributed in the sky, which contrasts with the elongated disc and bar shaped 
structures of the SMC main body.

\begin{figure}
\includegraphics[width=\columnwidth]{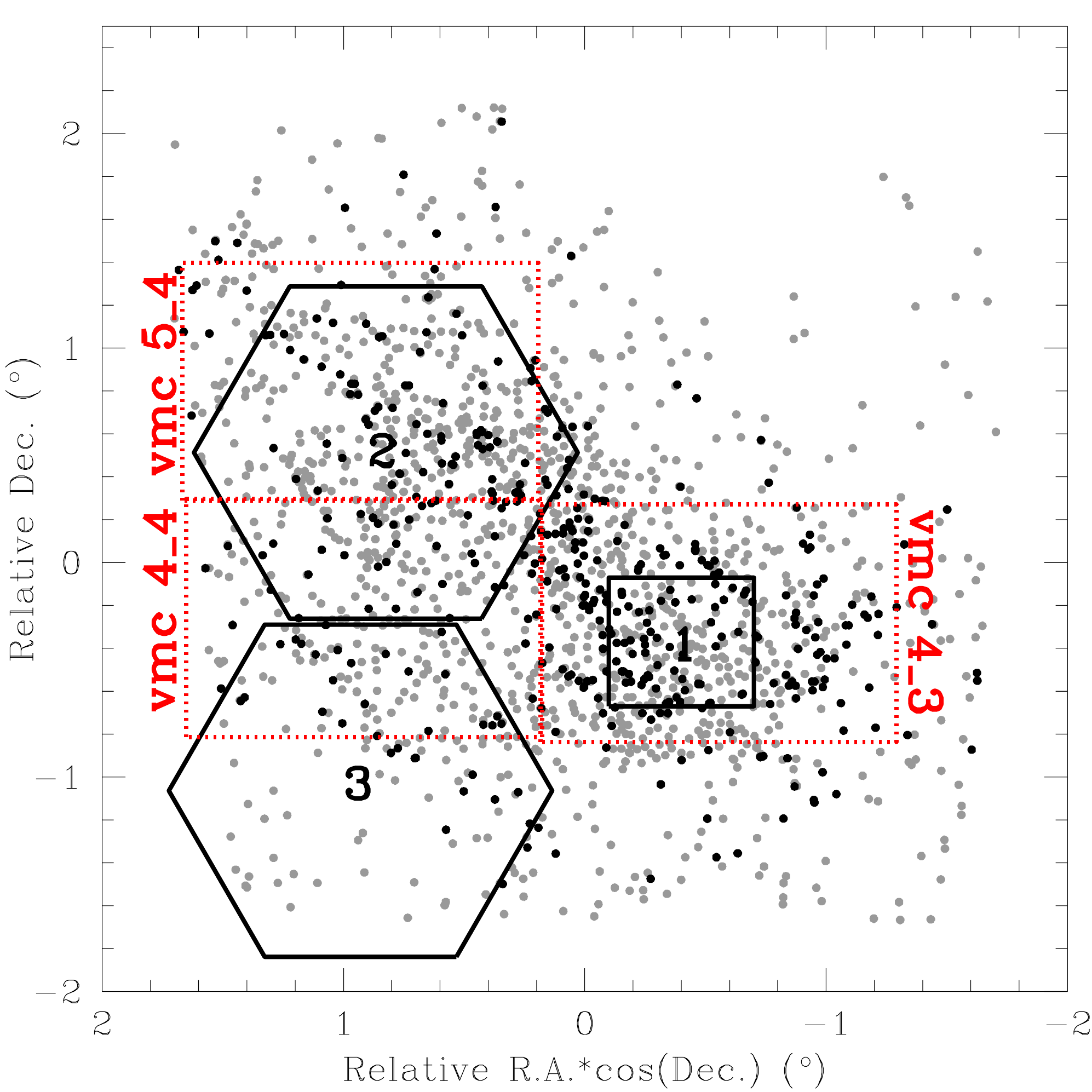}
    \caption{Spatial distribution of B18's objects (grey dots) and those of Table~\ref{tab:table2} 
( black dots) with the SMC fields studied by \citet{retal15} (red rectangles),
\citet{petal2016} (black rectangle) and by \citet{p17a} (hexagons) overplotted.}
   \label{fig:fig1}
\end{figure}

In this paper we show that the compilation of clusters built by B18 would not seem 
representative of the SMC  cluster population, so that by using it in statistical analyses
could not lead to meaningful results. Our approach is threefold: in Section 2, we firstly
conclude on the significant percentage of known relatively bright clusters that were
not included in B18. Then, we show that  most of the new detections could not be 
related to relative faint still unidentified clusters. In Section 3, we finally show that the 
formation history of those new detections does not match that of the real SMC clusters.
A summary of this work is presented in Section 4.

\section{The bulk of the SMC cluster population}

B18 matched their identified clusters with those in B08 for the same surveyed area,
and found that only 211 out of 515 clusters
were recovered at the end of their search-age estimate process. Their compilation
represent $\sim$ 40 per cent of B08's clusters.
At a first glance, this seems a relatively low percentage of cluster recovery as to
claim for accuracy and completeness in clusters detection. The matched clusters are included 
in their Table\,1 with the running ID number used by B08. Unfortunately, they did not include 
the various cluster names more frequently used. As for the 1108 new detections, B18 simply
did not find them in B08. However, after the cataloguing work by B08, more 
clusters have been identified and some few catalogued objects have been confirmed as possible
non-physical systems throughout the B18's covered area 
\citep[see e.g.,][]{getal10,pb12,petal2016}. 

We thus decided to build a catalogue as comprehensively as possible of SMC clusters
distributed throughout the B18's area with
age estimates derived from isochrone  fitting,
in order to perform a thorough comparison with  the B18's one. For clusters younger than 1 Gyr, we used two 
major catalogues constructed by \citet{chetal06} and \citet{getal10}. This age cut off was imposed
 because of 
the very well-known limiting magnitude of the photometric data
sets used by them, i.e, OGLE and MCPS, respectively (see figure 3 in B18). We complemented this gathering 
with ages derived by
\citet{petal07d,pietal08} and \citet{metal14} from slightly deeper Washington photometry.
For older clusters, we used a series of works based on Washington photometry \citep{p11b,p11c}. 
Whenever more than one age estimate is available, we averaged all values. Table~\ref{tab:table1}
shows the different literature sources that contribute to the resulting catalogue - distinguishing
the number of clusters found in B08 -, while the resulting list of the
419 clusters with the compiled averaged ages, their respective standard dispersion and the number of age 
values used are included in Table~\ref{tab:table2}. The B08's catalogue also comprises 139 clusters without
age estimates, so that our compilation of ages represents $\sim$ 75 per cent of the total 
number of clusters included in B08, \citet{chetal06} and \citet{getal10}, located within the
B18's area. Fig.~\ref{fig:fig1} shows the spatial distributions of B18's objects and those of
Table~\ref{tab:table2}, for comparison purposes.

\begin{table}
\caption{Literature sources used to build the catalogue of SMC clusters with age estimates.}
\label{tab:table1}
\begin{tabular}{@{}lcc}\hline
Reference           & Data set   & Number of clusters \\
\hline 
\citet{chetal06}    & OGLE       &  122 (B08) + 112   \\
\citet{getal10}     & MCPS       &  141 (B08) + 153\\
\citet{metal14}     & Washington &   29 (B08)  \\
\citet{petal07d}     & Washington &    5 (B08) \\
\citet{pietal08}     & Washington &    7 (B08) \\ 
\citet{p11c}        & Washington &    6 (B08)  \\
\citet{p11b}        & Washington &    9 (B08)  \\
\citet{pb12}        & Washington &    3 (B08) \\ 
\hline
\end{tabular}
\end{table}

\begin{table}
\caption{Age estimates of SMC clusters$^a$}
\label{tab:table2}
\begin{tabular}{@{}lccc}\hline
Cluster name & log(age) & $\sigma$(log(age)) & n \\\hline 
-- & -- & -- & -- \\
B55,SOGLE60 & 8.15 & 0.18 &3\\ 
B54,SOGLE62 & 8.20 & 0.10 &2 \\
H86-106w    & 8.67 & 0.03 &2 \\
-- & -- & -- & -- \\
\hline
\end{tabular}

\noindent $^a$ A portion of the table is presented here for
guidance of its contents. The entire table is available as
Supplementary material in the online version of the journal.
\end{table}

We then cross-correlated the B18's list of clusters with that of B08
on the basis of their coordinates using the IRAF\footnote{IRAF is distributed by  the 
National Optical Astronomy Observatories, 
which is operated by the Association of Universities for Research in Astronomy, Inc., 
under contract with the National Science Foundation.}.{\sc tmatch} task. It provides two tables 
that include matched and unmatched objects, respectively, within a certain tolerance distance.
We started with a distance of 0.60 arcmin to avoid multiple matchings. From the unmatched objects,
we run again {\sc tmatch} for a distance of 0.85 arcmin, and iterated the procedure from the
resulting unmatched sample for a distance of 1.00 arcmin. None execution of  {\sc tmatch} 
produced multiple matchings. We thus found a total of 159 clusters cross-identified. Note that
we performed the matching in a relatively relaxed fashion, because the tolerance distances employed are
bigger than the smallest SMC clusters, which are typically of $\sim$ 0.15-0.20 arcmin 
wide in radius \citep{petal2016}. Other additional 52 clusters were cross-correlated using distances 
between 1.00 and 4.00 arcmin. The matching of B18's compilation with \citet{chetal06}'s and \citet{getal10}
produced 11 (5\%) and 60 (20\%) identifications, respectively, which are very low numbers 
compared with those of other automatic cluster searching techniques 
\citep[][and references therein]{pieetal98,ivanovetal2017}. These outcomes reveal that the B18's compilation
of SMC clusters is not statistically complete (see also Section 3).

\begin{figure}
\includegraphics[width=\columnwidth]{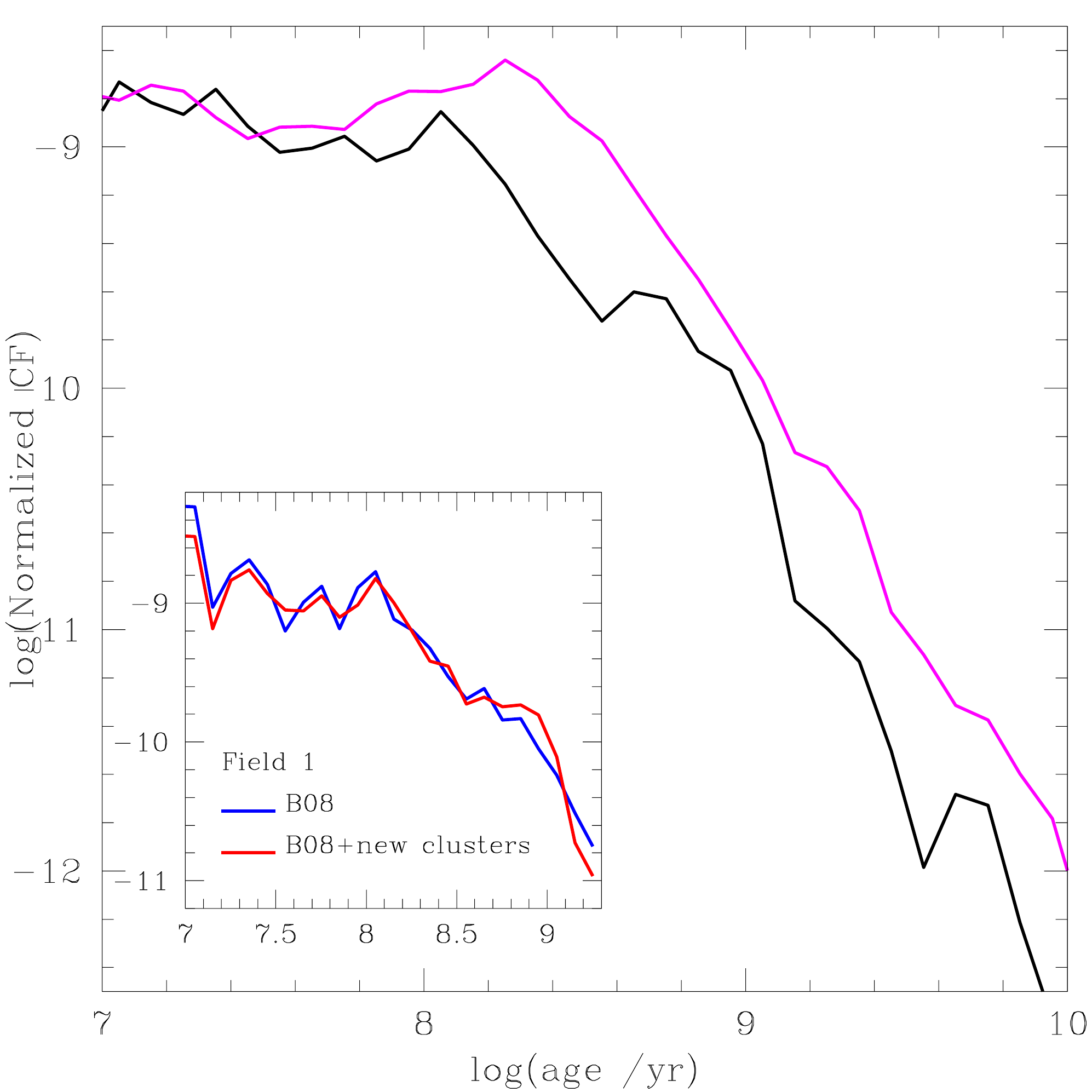}
    \caption{CFs for SMC clusters and B18's new detections drawn with black and magenta 
lines, respectively.
The inset panel depicts the CFs for two different cluster samples in Field\,1 
(see also Fig.~\ref{fig:fig1}).}
   \label{fig:fig2}
\end{figure}

Despite many relatively bright aggregates are not included in B18's Table\,1, their new detections
could be faint clusters, though. In order to probe this possibility, we took
advantage of two recent results on search of new SMC clusters. \citet{petal2016} used the 
VISTA\footnote{Infrared  Survey Telescope for Astronomy} near-infrared $YJK_s$  Magellanic Clouds 
survey data sets \citep[VMC,][]{cetal11} to explore a 36$\arcmin$$\times$36$\arcmin$ region centred on 
the South-west of the SMC bar (Field 1 in Table~\ref{tab:table3} and Fig.~\ref{fig:fig1}). 
The area is one of the most 
crowded and highly affected by
interstellar reddening in the galaxy, so that hidden clusters could exist there. They surveyed
that region looking for clusters with dimensions and mean stellar densities in the same ranges
than those previously known from optical passband photometric studies. They finally found
38 bona fide clusters, which represent an increase of $\sim$ 55  per cent in the number of known clusters
in that area. On the other hand, \citet{p17a} employed an upgraded version of the cluster searching technique 
developed by \citet{petal2016} to seek new clusters in a vast area of the SMC (Fields 2 and 3 
in Table~\ref{tab:table3} and Fig.~\ref{fig:fig1}). He made use of 
deep images from the Magellanic Stellar History (SMASH) survey and found very few new cluster 
candidates. Fig.~\ref{fig:fig1} illustrates  the position of the aforementioned fields.

We applied the {\sc tmatch} routine to cross-correlate the B18's catalogue with those built
by \citet{petal2016} and \citet{p17a}, respectively, following the recipe described above. 
We started with a tolerance distance  of 0.60 arcmin and increased it in steps of 0.15 arcmin up to 
1.00 arcmin; no multiple cross-correlations were produced. The results are shown in
Table~\ref{tab:table3}, where we successively listed for Fields 1, 2 and 3 the number of clusters 
already included in B08 and those recently discovered by \citet{petal2016} and \citet{p17a}, for
comparison purposes. The total number of B18's objects has been split in three columns, namely:
B08's clusters,  \citet{petal2016}'s and \citet{p17a}'s new clusters, and B18's new detections.
As can be seen, when comparing the different catalogues with that of B18,
we confirm the relatively high percentage of unmatched B08's clusters,
that goes from 44 per cent up to 59 per cent, with an average of 53 per cent. Unfortunately, a similar
conclusion is drawn for recently discovered  clusters; the number of new detections 
remaining notably high as well. As far as we are aware, the latter
would not appear to be related to genuine physical systems. Since B18's catalogue contains $\sim$ 84 per
cent of them, their interpretations of the SMC cluster formation history and that of SMC-LMC interaction 
should be considered with caution.

\begin{table}
\caption{Statistics of clusters in SMC fields.}
\label{tab:table3}
\begin{tabular}{@{}ccccccc}\hline
Field ID & B08          & New          & Ref. &       \multicolumn{3}{c}{B18's objects}  \\
     &       & clusters     &      &     B08     & New     & New  \\
         &                &              &      &             & clusters& detections\\
\hline 
1 &    68 & 38 & 1  & 29  & 6 & 81 \\
2 &   113 & 3  & 2  & 63  & 0 &  308 \\
3 &    34 & 3  & 2  & 14  & 0 & 110\\ 

\hline
\end{tabular}
\noindent Ref.: (1) \citet{petal2016}; (2) \citet{p17a}.
\end{table}

\section{The SMC cluster frequency}

Beside the statistics of clusters that provided us with a strong evidence on the content of the B18's
list of objects, the ages derived by them also play an important independent role in describing their
characteristics.  These age estimates tell us about the formation history
of the catalogued objects, so that in case they were real clusters, they should reproduce the formation 
history of the SMC cluster population.  Otherwise, the formation history constructed from their assigned 
ages will differ from that one.

\newpage

\begin{table*}
\small
\caption{Fitted coefficients for the CF/CF$_{\rm SFR}$ relationship.}
\label{tab:table4}
\begin{tabular}{@{}ccccccccccccc}\hline
Field & \multicolumn{4}{c}{VMC\,4$\_$3} & \multicolumn{4}{c}{VMC\,4$\_$4} & \multicolumn{4}{c}{VMC\,5$\_$4} \\
     & zero  & linear  &  rms & $\chi$$^2$ &  zero  & linear  & rms   & $\chi$$^2$ &  zero  & linear  & rms   & $\chi$$^2$ \\
             & point & term    &   &              & point & term    &   &            &    point & term    &   &            \\\hline

1   & 9.30  $\pm$ 0.71 & -1.17 $\pm$ 0.06 & 0.12 & 0.018 & 8.10 $\pm$ 1.40 & -1.01 $\pm$ 0.16 & 0.17 & 0.035 &   ---           &  ---         & ---  & ---  \\ 
2   & 10.28 $\pm$ 0.51 & -1.29 $\pm$ 0.06 & 0.07 & 0.006 & 6.20 $\pm$ 0.97 & -0.76 $\pm$ 0.10 & 0.12 & 0.018 & 3.62 $\pm$ 0.61 & -0.48 $\pm$ 0.07 & 0.04 & 0.002\\   
3   & 10.55 $\pm$ 0.62 & -1.35 $\pm$ 0.07 & 0.05 & 0.003 & 7.13 $\pm$ 1.53 & -0.93 $\pm$ 0.17 & 0.26 & 0.075 & 7.03 $\pm$ 1.45 & -0.92 $\pm$ 0.17 & 0.24 & 0.065 \\    
4   & 7.45  $\pm$ 2.77 & -0.94 $\pm$ 0.31 & 0.22 & 0.060 & 5.17 $\pm$ 1.01 & -0.64 $\pm$ 0.11 & 0.07 & 0.005 & 6.90 $\pm$ 1.54 & -0.90 $\pm$ 0.18 & 0.15 & 0.027\\ 
5   & 12.18 $\pm$ 1.11 & -1.51 $\pm$ 0.13 & 0.12 & 0.018 &10.21 $\pm$ 1.00 & -1.22 $\pm$ 0.19 & 0.24 & 0.050 &    ---        &  ---          & --- & ---  \\
6   & 8.00  $\pm$ 1.00 & -1.00 $\pm$ 0.11 & 0.17 & 0.032 &10.42 $\pm$ 0.95 & -1.31 $\pm$ 0.11 & 0.25 & 0.070 & 5.27 $\pm$ 0.59 & -0.75 $\pm$ 0.07 & 0.18 & 0.036\\ 
7   & 4.81  $\pm$ 0.42 & -0.62 $\pm$ 0.04 & 0.08 & 0.008 & 9.56 $\pm$ 1.64 & -1.20 $\pm$ 0.19 & 0.16 & 0.031 & 5.19 $\pm$ 0.49 & -0.72 $\pm$ 0.06 & 0.16 & 0.028 \\ 
8   & 10.96 $\pm$ 1.27 & -1.32 $\pm$ 0.14 & 0.10 & 0.012 & 9.23 $\pm$ 0.94 & -1.15 $\pm$ 0.10 & 0.14 & 0.024 & 6.27 $\pm$ 0.94 & -0.82 $\pm$ 0.11 & 0.13 & 0.021 \\ 
9   & 6.28  $\pm$ 1.28 & -0.86 $\pm$ 0.15 & 0.27 & 0.080 &     ---         &  ---             & ---  & ---   & 5.00 $\pm$ 0.86 & -0.62 $\pm$ 0.10 & 0.09 & 0.010 \\ 
10  & 12.21 $\pm$ 1.83 & -1.49 $\pm$ 0.21 & 0.15 & 0.026 & 8.92 $\pm$ 1.59 & -1.11 $\pm$ 0.14 & 0.18 & 0.039 & 6.48 $\pm$ 1.48 & -0.80 $\pm$ 0.17 & 0.10 & 0.013 \\   
11  & 9.80  $\pm$ 3.36 & -1.20 $\pm$ 0.39 & 0.09 & 0.013 & 7.86 $\pm$ 0.54 & -1.00 $\pm$ 0.06 & 0.12 & 0.016 & 6.12 $\pm$ 1.78 & -0.77 $\pm$ 0.23 & 0.12 & 0.019 \\ 
12  & 8.50  $\pm$ 3.90 & -1.00 $\pm$ 0.44 & 0.12 & 0.020 &10.89 $\pm$ 1.65 & -1.39 $\pm$ 0.19 & 0.19 & 0.041 &    ---         &  ---        & ---  & ---  \\
\hline
\end{tabular}
\end{table*}

It has been shown that the so-called cluster frequency (CF) is a more appropriate tool to
describe the cluster formation history than the age histograms \citep[see,e,g,][]{baetal13,p14b,piskunovetal2018}. 
The former traces the number of clusters
per time unit, while histogram bins could span different time interval. For instance, a histogram in
log(age) with bin sizes of 0.1 embraces periods of time of $\sim$ 2.6 Myr and $\sim$ 260 Myr at the age intervals of
log(age) = 7.0-7.1 and 9.0-9.1, respectively. This uneven split of the whole time period
could produce spurious peaks in the number of clusters at certain age bins, thus misleading the 
interpretation about enhanced periods of cluster formation, periods of more intense cluster dissolution, etc.,
among others. Particularly,  the analysis of the SMC cluster formation history and the 
interaction with the LMC carried out B18 relies on age histograms.

In order to build the SMC CF we considered each age estimate of Table\,2 as represented by an 
one-dimensional {\it Gaussian} of unity area centred at the respective age value, with a FWHM/2 equals to the age error. 
Then, we used a grid of age bins with sizes of $\Delta$(log(age)) = 0.1  and added the fractions of the {\it
Gaussian}s' areas that fall into the bin boundaries. For instance, a point that is centred at any age
interval and has an error smaller than $\Delta$(log(age)) = 0.05 contributes with an amount of 1.0 to the total
number of clusters to that age bin. Thus, by taking into account the uncertainties of the age estimates, we
were able  to produce an intrinsic SMC cluster age distribution. We then divided the total number of clusters 
per age bin by 
the size of the respective interval to obtain the corresponding CF. Fig.~\ref{fig:fig2} shows the resulting CF
drawn with a solid black line, normalised to the total number of clusters for comparison purposes. 

Although
it has been built with $\sim$ 75 per cent of all the catalogued clusters located within the B18's area, it 
results statistically representative of the whole SMC cluster population in that region. 
Indeed, \citet{mk2011}  and \citet{p14b}
have shown that cut-offs of the low-mass cluster regime (cluster mass $\lesssim$ 10$^3$ $M_{\odot}$) 
does not impact in the shape of the resulting CF. With a mass cut-off at 5$\times$10$^3$  $M_{\odot}$, 
accurate CFs are still feasible to obtain.  Nevertheless, for the sake of the reader we 
separately built the CFs for 
clusters located in Field\,1, using the previously known 68 ones (B08) and the 
new discovered 38 clusters  by \cite{petal2016}
(see Table~\ref{tab:table3}), respectively. The inset panel of Fig.~\ref{fig:fig2} presents both resulting CFs. 
As can be seen, the new relatively faint clusters are imprinted with the same formation history than the
most massive ones. Therefore, if the B18's new detections were faint genuine clusters, their CF should look
like the one we built for the SMC. However, by following the same procedure described above to build the SMC CF, 
we constructed that from the B18's new detections, which resulted remarkably different to the former, as 
judged by the magenta solid line drawn in Fig.~\ref{fig:fig2}. Nevertheless, the B18's sample does not
include relatively faint objects, because of the limiting magnitude of the photometric data sets used 
(see discussion in Section 2).

\subsection{Cluster dissolution across the SMC}

The CF constructed from clusters compiled in Table~\ref{tab:table2} represent the overall
present-day distribution of the SMC cluster population as a function of age. However, 
such a distribution could vary with the position in the galaxy. For instance, \citet{p14b} 
used the \citet{hz09}'s LMC regions to show that there exist some variations of the CFs 
from one region to another: 30 Doradus is the region with the highest relative CF for the
youngest clusters, while the inner LMC regions have larger numbers of clusters
during the period 1-3 Gyr than the outer ones. On the other hand, \citet{baetal13}
showed that the LMC cluster dissolution has played a role for clusters older than 200 Myr.
By adopting the star formation rate (SFR) derived by \citet{hz09} as the cluster formation
rate (CFR), they found that about 90 per cent of them are lost per dex of lifetime.
Therefore, in general, the present-day CF could result in a complex function which depends 
on both the local CFR and the cluster disruption.

\begin{figure*}
\includegraphics[width=\textwidth]{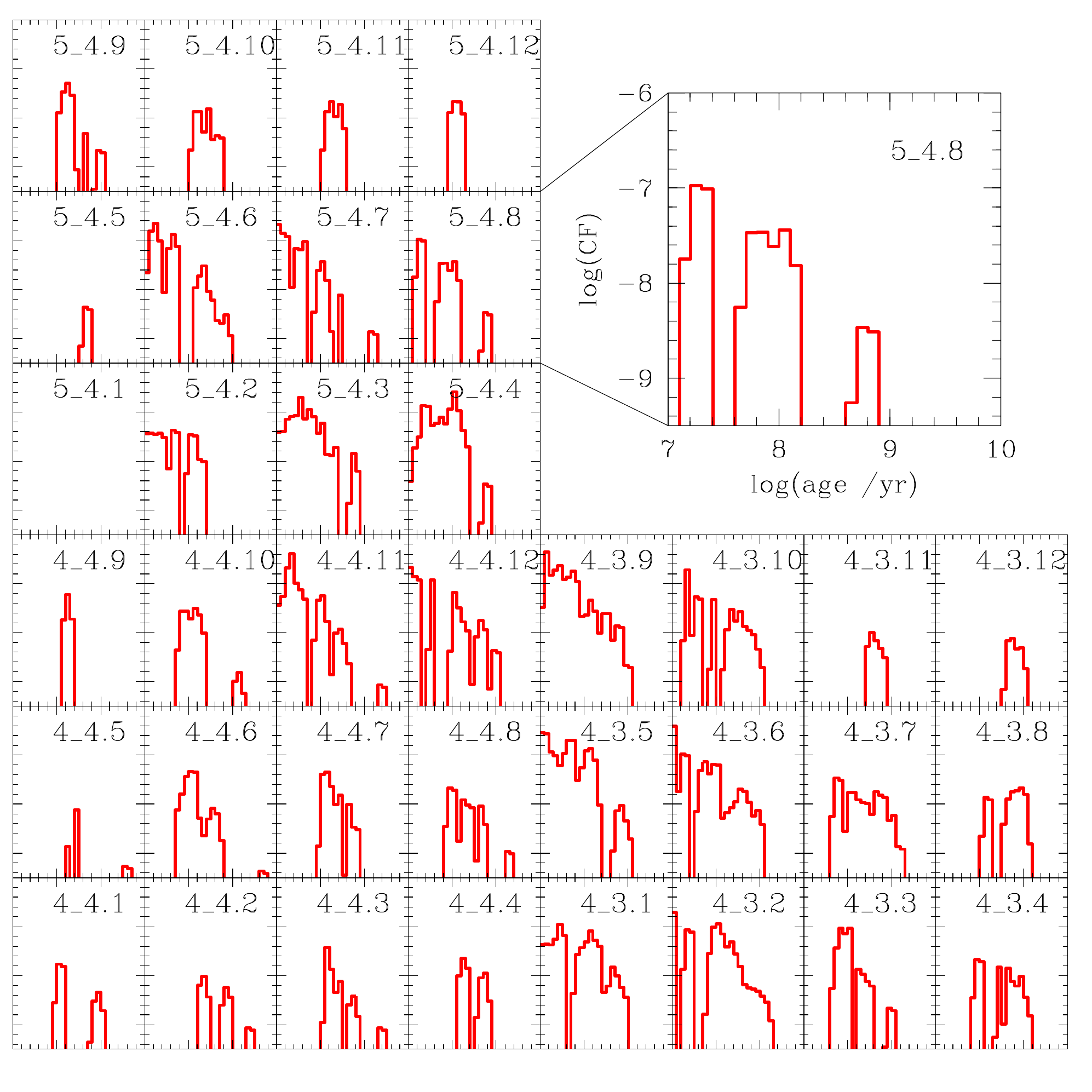}
    \caption{CFs for different SMC sub-fields placed following their spatial distribution pattern (see 
Fig.~\ref{fig:fig1} and details in Section 3.1). Each panel has the same axes as illustrated
in the top-right panel.}
   \label{fig:fig3}
\end{figure*}

\citet{retal15} made use of the VMC database to build the SMC SFR
producing not only an overall SFR for the whole galaxy, but also local SFRs for homogeneously
distributed regions of $\sim$ 22$\arcmin$$\times$22$\arcmin$ (see their figure 5). Three 
VMC tiles span over the SMC main body as illustrated in Fig.~\ref{fig:fig1}, so that
we take advantage here of the respective SFRs. \citet{retal15} split each tile
in twelve sub-fields: four columns along the Right Ascension axis 
and three rows along the Declination axis. The sub-fields, from $\#$1 to  $\#$12 were allocated
across each tile from the bottom-left corner to the top-right one, i.e., by decreasing Right Ascension 
and increasing Declination. We used those SFRs to build CFs by adopting a power-law cluster
mass distribution with a slope $\alpha$ = -2 and assuming that clusters and field stars
share the same formation rates (SFR $\equiv$ CFR). CFs and CFRs are liked through the expression:

\begin{equation}
CF_{SFR} = SFR \times \frac{\Sigma\, m^{-2}}{\Sigma\, m^{-1}} 
\end{equation}

\noindent where $m$ is the cluster mass and the sums are computed over the SMC cluster mass range. 
Additionaly, we constructed CFs for each sub-field from our compilation of clusters with age
estimates (Table~\ref{tab:table2}) following the same recipe employed above, i.e., by considering
each age value as a {\it Gaussian} of unity area and adding the contribution of each {\it Gaussian}
to different age intervals. The resulting present-day sub-fields CFs are depicted in Fig.~\ref{fig:fig3}.

\begin{figure*}
\includegraphics[width=\textwidth]{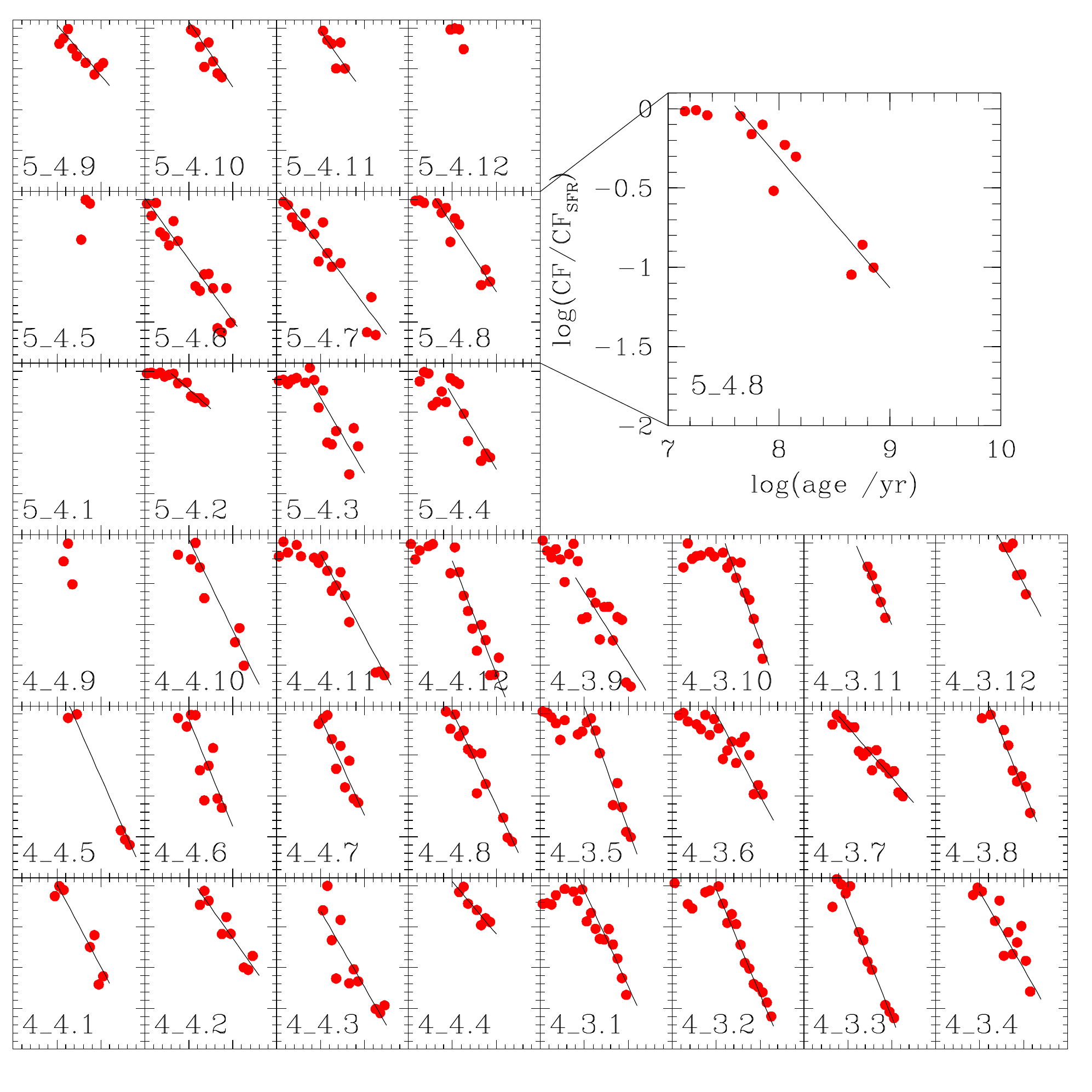}
    \caption{CF to CF$_{SFR}$ ratios for different SMC fields placed following their spatial distribution 
pattern (see Fig.~\ref{fig:fig1} and details in Section 3.1). Each panel has the same 
axes as illustrated in the top-right panel. The straight lines represent the fitted
linear relationships (see also Table~\ref{tab:table4}).}
   \label{fig:fig4}
\end{figure*}

A close inspection of Fig.~\ref{fig:fig3} reveals that the cluster formation has not been
continuous, but a process with relative short and long periods that has varied with the
position in the galaxy. At the same time, it is possible to connect sub-fields that have
experienced cluster formation at the same time, namely: 1) relatively old formation episodes
in the outermost Western sub-fields (VMC\,4$\_$3.8 and 4$\_$3.12); 2) no recent (log(age) $<$ 8) 
cluster formation in outer sub-fields (VMC\,5$\_$4.9, 5$\_$4.10, 5$\_$4.11, 5$\_$4.12, 4$\_$4.2, 
4$\_$4.3, 4$\_$4.4 and 4$\_$3.11) and; 3) intense cluster formation in the SMC bar (VMC\,5$\_$4.3, 5$\_$4.4, 
4$\_$4.12, 4$\_$3.1, 4$\_$3.2, 4$\_$3.5, 4$\_$3.6, 4$\_$3.9). All these features are compatible
with an outside-in formation process \citep[see, e.g.][]{netal09,setal09,wetal13}. On the other hand,
isolated relatively short cluster enhancements are seen towards the outermost Eastern sub-fields 
(VMC\,4$\_$4.1, 4$\_$4.5, and 4$\_$4.9), which could be part of the onset of the Magellanic bridge 
where recent field star and cluster formation took place \citep{petal15a,bicaetal2015,mackeyetal2017}.

We divided the present-day sub-field CFs by the respective CF$_{\rm SFR}$ obtained from eq. (1)
in order to find out any hint of dissolution. The results are presented in Fig.~\ref{fig:fig4}. 
Similarly to \citet{baetal13}'s outcomes, the
number of observed clusters and those estimated from the \citet{retal15}'s SFR are comparable for 
ages younger than log(age) $\sim$ 7.8 - 8.2, depending on the position in the galaxy. For older ages, 
the number of observed clusters clearly turn to decrease as a consequence of their dissolution with 
time. We fitted linear relationships of the CF/CF$_{\rm SFR}$ ratios as a function of log(age) for the period
where cluster dissolution is detected. The resulting coefficients are shown in Table~\ref{tab:table4},
while Fig.~\ref{fig:fig4} illustrates the derived linear relationships. 

\begin{figure}
\includegraphics[width=\columnwidth]{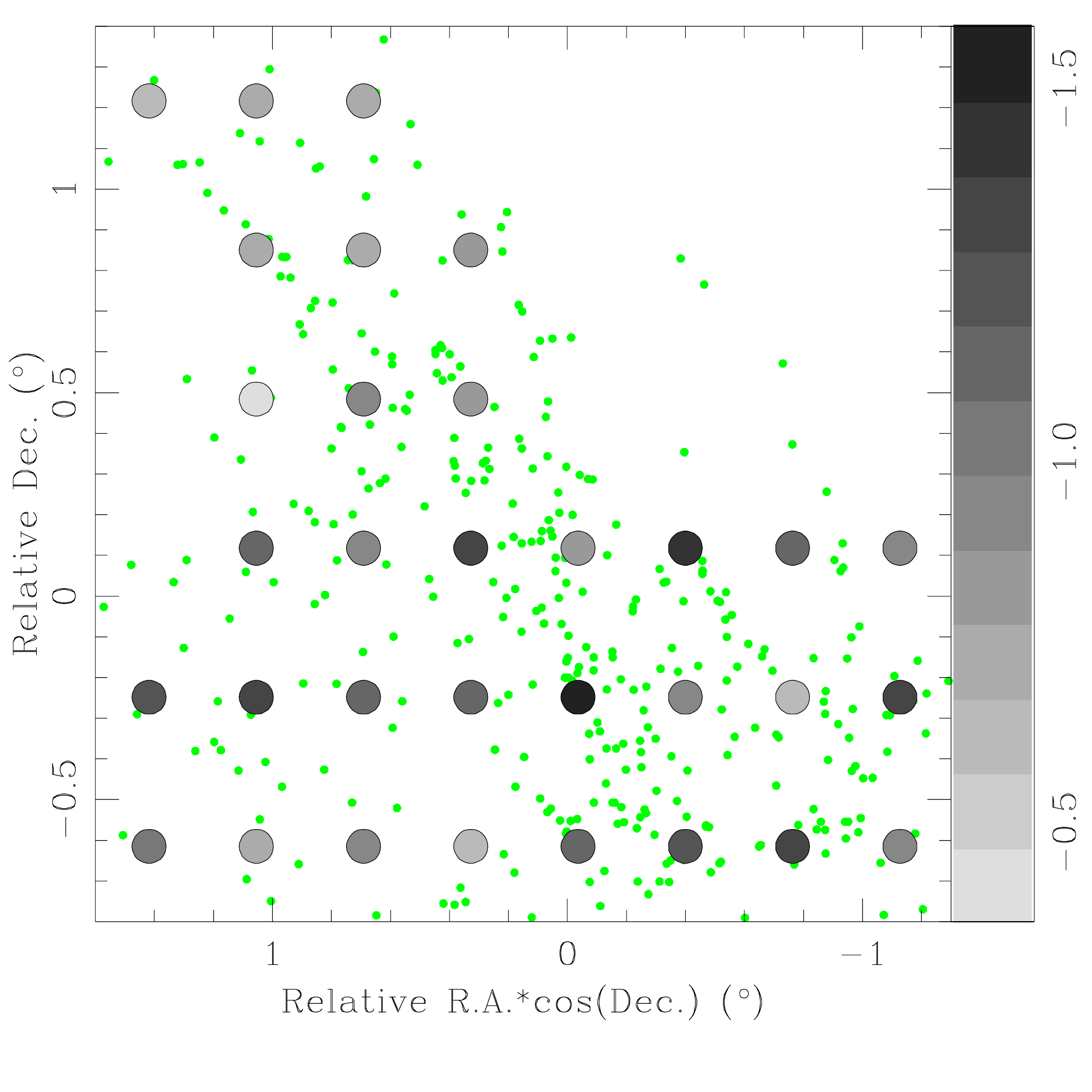}
    \caption{Spatial distribution of clusters in Table~\ref{tab:table2} (green dots).
We superimposed the CF/CF$_{SFR}$ ratio slopes (Table~\ref{tab:table4}) represented by 
grey-scale filled circles -according the colour bar- for different SMC fields.}
   \label{fig:fig5}
\end{figure}

As can be seen, 
the slope - which is a measure of the dissolution rate - changes with the position in the galaxy.
In order to highlight such a variation, we produced Fig.~\ref{fig:fig5}, which shows the spatial
distribution of SMC clusters with green dots and grey-scale filled circles placed at the centre of the 
VMC sub-fields representing the respective CF/CF$_{\rm SFR}$ slope values. Sub-fields in the Northern 
part of the SMC main body ($\Delta$(Dec.) $>$ 0.5$\degr$) seem to share similar relatively low disruption rates, 
while those of the innermost part of the SMC, as well as those at the onset of the Magellanic bridge, have 
higher rates of dissolution. Since cluster disruption is mainly caused by the SMC tidal field, we speculate with the
possibility that these findings confirm that clusters located in more massive galactic regions
suffer more pronounced tidal effects. In addition, the relative large CF/CF$_{\rm SFR}$ slope values
for sub-fields located towards the South-East could be caused by the interaction with the LMC, which has been
detaching the SMC gas and stellar content along the Magellanic bridge 
\citep{setal14,noeletal2015,carreraetal2017}. Notice that if the cluster formation were
not assumed to be that of the respective star field, the results of Fig.~\ref{fig:fig5} 
would reveal that the CFR could differ significantly with respect to the SFR with position 
in the galaxy.

\section{Conclusions}

In this study we addressed the issue of the unexpected large number of
new cluster candidates compiled by B18. These candidates represent more
than two hundred per cent increase of the known SMC cluster population
%, but
%also their whole sample is much larger than that we would {\bf expect,} considering
%the galaxy mass in comparison with that of the LMC and the size of its 
%population of clusters.

We showed that the fully-automatic procedure implemented by B18 to detect and
derive age estimates of cluster candidates did not completely recover the previously
known bright catalogued clusters, neither those relatively faint ones recently
discovered. To arrive to such a conclusion, we  thoroughly searched the literature
in order to compile a list of clusters as comprehensive as possible,
including those listed in major cataloguing works and those from
most recent cluster searches in particular SMC regions.
We then used a proven technique that reliably cross-identified each object in  
the B18's catalogue to its most probable counterpart in our own compilation
and found that, in average, more than 50 per cent of the known clusters 
located in the SMC main body are not included in B18. Notably, 
a huge amount of B18's new detections remained unmatched.

Nearly 75 per cent of all catalogued clusters have previous age estimates
based on isochrone fitting to the cluster CMDs. After averaging the
individual age values found in the literature, we constructed their global
CF, which represents the intrinsic distribution of SMC clusters per time unit 
as a function of time. To build such a CF, we took into account the age
uncertainties, so that the resulting distribution does not depend on the
chosen age bins. Likewise, we showed that its shape would not vary if we
used the entire SMC cluster population. We also produced a CF for the B18's 
new detections  ($\sim$ 84 per cent of their whole sample) and found that it 
turned out to be noticeably different from the SMC CF. Hence, we concluded that 
the B18's catalogued could be contaminated with non-cluster objects.

We finally analysed the dependence of the CF with the position in the
galaxy. In order to do that, we built CFs for regions of 
$\sim$ 0.13 square degrees homogeneously distributed across the SMC
main body. The resulting CFs confirm the outside-in cluster formation
scenario, the recent vigorous formation activity that took place in the bar, 
as well as that in the onset of the Magellanic bridge. By assuming that
clusters and field stars share the same formation history, we computed
the number of clusters formed per time unit using a spatially resolved
SFR. We found that the cluster dissolution rate over time varies
with the position in the galaxy. The Northern part of the SMC main body
is pictured by relatively small dissolution rate values in comparison
with those of the SMC bar, and even those of the onset of the
Magellanic bridge, where cluster disruption would seem to have been more 
important. We thus provide, for the first time, an observational
evidence in the sense that the stronger a galactic gravitational
field (e.g., larger concentration of galaxy mass, tidal forces from interaction
with the LMC), the higher the cluster dissolution rate.

\section*{Acknowledgements}

%This publication makes use of data products from the Two Micron All Sky Survey, 
%which is a joint project of the University of Massachusetts and the Infrared 
%Processing and Analysis Center/California Institute of Technology, funded by 
%the National Aeronautics and Space Administration and the National Science Foundation.
 We thank the referee for the thorough reading of the manuscript and
timely suggestions to improve it. 

%%%%%%%%%%%%%%%%%%%%%%%%%%%%%%%%%%%%%%%%%%%%%%%%%%



\bibliographystyle{mnras}
%\bibliography{paper} % if your bibtex file is called paper.bib

%to be uncommented before sending to editor
\input{paper.bbl}

%\begin{thebibliography}{99}
%\end{thebibliography}

%%%%%%%%%%%%%%%%%%%%%%%%%%%%%%%%%%%%%%%%%%%%%%%%%%

%%%%%%%%%%%%%%%%% APPENDICES %%%%%%%%%%%%%%%%%%%%%

%\appendix


%%%%%%%%%%%%%%%%%%%%%%%%%%%%%%%%%%%%%%%%%%%%%%%%%%

% Don't change these lines
\bsp	% typesetting comment
\label{lastpage}
\end{document}